\documentclass[11pt]{article}
\usepackage{amsmath,amssymb,amsfonts,graphicx,dsfont}
\usepackage{graphicx}
\usepackage{amsfonts}
\usepackage{dsfont}
\usepackage{amssymb}
\usepackage{amsmath}
\usepackage{amsthm}
\usepackage{epsf}
\usepackage{graphicx,psfrag,color,pstcol,pst-grad}
\usepackage{latexsym}
\usepackage{calc}
\usepackage{multicol}
\usepackage{subfigure}
\usepackage{pst-tree}
\usepackage{dsfont}
\usepackage{amsthm}
\usepackage{rotate}

\DeclareMathOperator{\card}{card}

\def\OMIT#1{}
\def\INC#1{#1}

\def\Xcal{\mathcal{X}}

\def\Fcal{\mathcal{F}}
\def\RR{\mathbb{R}}
\newcommand {\br}[1]{\left(#1\right)}
\def\PP{P}
\def\KK{K}

\title{A covariance kernel for protein sequences}
\author{
Marco Cuturi\\
Computational Biology Group\\
Ecole des Mines de Paris \\
35 rue Saint Honor\'e
\\ 77300 Fontainebleau\\
\texttt{marco.cuturi@ensmp.fr} \and
Jean-Philippe Vert\\
Computational Biology Group\\
Ecole des Mines de Paris \\
35 rue Saint Honor\'e
\\ 77300 Fontainebleau\\
\texttt{jean-philippe.vert@ensmp.fr} }

\begin{document}
\date{}
\maketitle

\begin{abstract}

We propose a new kernel for biological sequences which borrows
ideas and techniques from information theory and data compression.
\INC{This kernel can be used in combination with any kernel
method, in particular Support Vector Machines for protein
classification.} By incorporating prior biological assumptions on
the \OMIT{structure}\INC{properties} of amino-acid sequences and
using a Bayesian \OMIT{mixture}\INC{averaging} framework, we
compute the value of this kernel in linear time and space\INC{,}
benefiting from previous achievements proposed \OMIT{by}\INC{in
the field of} universal coding. \INC{Encouraging
}\OMIT{C}\INC{c}lassification results are reported on a standard
protein homology detection experiment.

\end{abstract}
\section{Introduction}
\par The need for efficient analysis and classification tools for protein sequences is more than ever a
core problem in computational biology. In particular, the
availability of an ever-increasing quantity of biological
sequences calls for efficient and computationally feasible
algorithms to detect functional similarities between DNA or
amino-acid sequences, cluster them, and annotate them.

\par Recent years have witnessed the rapid development of a class of algorithms called {\em kernel methods} \cite{schoelkopf02learning} that may offer useful tools for these tasks. In particular, the Support Vector Machine (SVM) algorithms \cite{bose92,vapnik98statistical} provide state-of-the-art performance in many real-world problems of classifying objects into predefined classes. SVMs have already been applied with success to a number of issues in computational biology, including but not limited to protein homology detection \cite{jaakola00discriminative,leslie02spectrum,leslie03mismatch,noble02combining,BenBru03,VerSaiAku04} functional classification of genes \cite{LiaNob02,vert-tree}, or prediction of gene localization ~\cite{HuaSun01b}. A more complete survey of the application of kernel methods in computational biology is presented in the forthcoming book \cite{schoelkopf04kernel}.

The basic ingredient shared by all kernel methods is the kernel
function, that measures similarities between pairs of objects to
be analyzed or classified, protein sequences in our case. While
early-days SVM focused on the classification of vector-valued
objects, for which kernels are well understood, recent attempts to
use SVM for the classification of more general objects have
resulted in the development of several kernels for strings
\cite{watkins00dynamic,haussler99convolution,jaakola00discriminative,leslie02spectrum,leslie03mismatch,noble02combining,BenBru03,VerSaiAku04},
graphs \cite{KasTsuIno03}, or even phylogenetic profiles
\cite{vert-tree}.

A useful kernel for protein sequences should have several
properties. It should be rapid to compute (typically, have a
linear complexity with respect to the lengths of the compared
sequences), represent a biologically relevant measure of
similarity, be general enough to be applied without tuning on
different datasets, yet efficient in terms of classification
accuracy. Such an ideal kernel probably does not exist, and
different kernels might be useful in different situations. For
large-scale or on-line applications, the computation cost becomes
critical and only fast kernels, such as the spectrum
\cite{leslie02spectrum} and mismatch \cite{leslie03mismatch}
kernels can be accepted. In applications where accuracy is more
important than speed, slower kernels that include more biological
knowledge, such as the Fisher \cite{jaakola00discriminative} or
local alignment \cite{VerSaiAku04} kernels might be accepted if
they improve the performance of a classifier.

Our contribution in this paper is to introduce a new class of
kernels for string that are both rapid to compute (they have a
linear-time complexity in time and memory), while still including
biological knowledge. The biological knowledge takes the form of a
family of probabilistic models for sequences supposed to be useful
to model general classes of proteins. The ones we consider are
variable-length Markov chains, also known as context-tree models
\cite{willems95contexttree} or probabilistic suffix trees
\cite{bejerano99modeling}. These models offer three advantages:
first, they have been shown to be useful to represent protein
families \cite{bejerano99modeling,smt}, second, they can have
different degrees of generality by varying the suffix-tree,
allowing then to model larger or smaller classes of sequences, and
third, their structure enables us to derive a kernel that can be
implemented in linear time and space with respect to the sequence
length. The last two features would not be shared by more complex
models such as hidden Markov models \cite{DurEddKroMit98}. A
second source of biological information is represented by a prior
distribution on the models, including the use of Dirichlet
mixtures \cite{DurEddKroMit98} to take into account similarities
between amino-acids.

As opposed to the classical use of probabilistic models to model
families of sequences \cite{bejerano99modeling,smt} or to the
Fisher kernel, we don't perform any parameter or model estimation.
We rather project each sequence to be compared to the set of all
distributions in the probabilistic models, and compare different
sequences through their respective projections. The resulting
kernel belongs to the class of covariance kernels introduced in
\cite{ seeger02covariance}. Formally, the computation of the
kernel boils down to computing some posterior distribution for
pairs of sequences in a Bayesian framework. The computation can be
performed efficiently thanks to a clever factorization of the
family of context-tree models using a trick presented in
\cite{willems95contexttree}. The resulting kernel can be
interpreted in the light of noiseless coding theory
\cite{CovTho91}: it is related to the gain in redundancy when the
two sequences compared are compressed together, and not
independently from one another.

The paper is organized as follows. In Section
\ref{sec:correlation} we present the general strategy of making
covariance kernels from families of probabilistic models. In
Section \ref{sec:ctw} we define a kernel for protein sequences
based on context-tree models. Its efficient implementation is
presented in Section \ref{sec:implementation}, and some of its
properties are discussed in Section \ref{sec:properties}.
Experimental results on a benchmark problem of remote homology
detection is presented in Section \ref{sec:experiment}

\section{Probabilistic models and covariance kernels}\label{sec:correlation}
\par A (parametric) probabilistic model on a measurable space $\Xcal$ is a family of distributions
$\{P_\theta,\theta\in\Theta\}$ on $\Xcal$, where $\theta$ is  the
parameter of the distribution $P_{\theta}$. Typically, the set of
parameters $\Theta$ is a subset of $\RR^n$, in which case $n$ is
called the dimension of the model. As an example, a hidden Markov
model (HMM) for sequences is a parametric model, the parameters
being the transition and emission probabilities
\cite{DurEddKroMit98}. A family of probabilistic models is a
family $\{P_{f,\theta_f}, f \in \Fcal, \theta_f\in\Theta_f\}$,
where $\Fcal$ is a finite or countable set, and $\Theta_f \subset
\RR^{\text{dim}(f)}$ for each $f \in \Fcal$, where $\text{dim}(f)$
denotes the dimension of $f$. An example of such a family would be
a set of HMMs with different architectures and numbers of states.
Probabilistic models are typically used to model sets of elements
$X_1,\ldots,X_n \in \Xcal$, by selecting a model $\hat{f}$ and a
choosing a parameter $\hat{\theta}_{\hat{f}}$ that best ''fits''
the dataset, using criteria such as penalized maximum likelihood
or maximum a posteriori probability \cite{DurEddKroMit98}.

Alternatively, probabilistic models can also be used to
characterize each single element $X \in \Xcal$ by the
representation $\phi(X) = \br{P_{f,\theta_f}(X)}_{f \in \Fcal,
\theta_f \in \Theta_f}$. If the probabilistic models are designed
in such a way that each distribution is roughly characteristic of
a class of objects of interest, then the representation $\phi(X)$
quantifies how $X$ fits each class. In this representation, each
distribution can be seen as a filter that extracts from $X$ an
information, namely the probability of $X$ under this
distribution, or equivalently how much $X$ fits the class modelled
by this distribution.

Kernels are real-valued function $K:\Xcal\times\Xcal \rightarrow
\RR$ that can be written in the form of a dot product
$K(X,Y)=\langle\psi(X),\psi(Y)\rangle$ for some mapping $\psi$
from $\Xcal$ to a Hilbert space \cite{schoelkopf02learning}. Given
the preceding mapping $\phi$, a natural way to derive a kernel
from a family of probabilistic models is to endow the set of
representations $\phi(X)$ with a dot product, and set
$K(X,Y)=\langle\phi(X),\phi(Y)\rangle$. This can be done for
example if a prior probability $\pi(f,d\theta_f)$ can be defined
on the set of distributions in the models, by considering the
following dot product:
\begin{equation}\label{eqn:k1}
K(X,Y) = \langle\phi(X),\phi(Y)\rangle \overset{\text{def}}{=}
\sum_{f \in \Fcal} \pi(f) \int_{\Theta_f}
P_{f,\theta_f}(X)P_{f,\theta_f}(Y) \pi(d\theta_f | f) .
\end{equation}
By construction, the kernel (\ref{eqn:k1}) is a valid kernel, that
belongs to the class of covariance kernels
\cite{seeger02covariance}. Observe that contrary to the Fisher
kernel that also uses probabilistic models to define kernel, no
model or parameter estimation is required in (\ref{eqn:k1}).
Intuitively, for any two elements $X$ and $Y$ the kernel
(\ref{eqn:k1}) automatically detects the models and parameters
that explain both $X$ and $Y$ well.

There is of course some arbitrary in this kernel, both in the
definition of the models and in the choice of the prior
distribution $\pi$. This arbitrary can be used to include prior
(biological) knowledge in the kernel. For example, if one wants to
detect similarity with respect to families of sequences known to
be adequately modelled by HMMs, then using HMM models constrains
the kernel to detect such similarities. We use this idea below to
define a set of models and prior distributions for protein
sequences.

As the probability of a sequence under the models we define below
decreases roughly exponentially with its lengths, the value of the
kernel (\ref{eqn:k1}) can be strongly biased by differences in
length between the sequences, and can take exponentially small
values. This is a classical issue with many string kernels that
leads to bad performance in classification with SVM
\cite{SchWesEskLesNob02,VerSaiAku04}. This undesirable effect can
easily be controlled in our case by normalizing the likelihoods as
follows:
\begin{equation}\label{eqn:k2}
K_\sigma(X,Y) = \sum_{f \in \Fcal} \pi(f) \int_{\Theta_f}
P_{f,\theta_f}(X)^{\frac{\sigma}{|X|}}P_{f,\theta_f}(Y)^{\frac{\sigma}{|Y|}}
\pi(d\theta_f | f) .
\end{equation}
where $\sigma$ is a width parameter and $|X|$ and $|Y|$ stand for
the lengths of both sequences. Equation (\ref{eqn:k2}) is clearly
a valid kernel (only the feature extractor $\phi$ is modified),
and the parameter $\sigma$ controls the range of values it takes.

\section{A covariance kernel based on context-tree models}\label{sec:ctw}
\par In this Section we derive explicitly a covariance kernel for
strings based on context-tree models with mixture of Dirichlet
priors.
Context-tree models are Markovian models which define an efficient
framework to describe constraints on amino-acid successions in proteins, as validated by their use in \cite{bejerano99modeling,smt}.
Dirichlet priors offer a biologically meaningful estimation of the
likelihood of such transitions by giving an a-priori knowledge on
the multinomial parameters which parameterize Markovian models
transitions.

\subsection{Framework and notations}
Starting with basic notations and definitions, let $E$ a
finite set of size $d$ called the alphabet. Practically speaking
$E$ can be thought of the 20 letters alphabet of amino-acids. For
a given depth $D\in\mathds{N}$ corresponding to the maximal memory
of our Markovian models we note
$\mathcal{X}=\cup_{n=0}^{\infty}(E^D \times E)^n$ the set on which
we define our kernel. Observe that we do not define directly the kernel on the set of finite-length sequences, but rather in a slightly more general framework more amenable to future generalization discussed in Section \ref{sec:sampling} below. $M$ is the set of strings of $E$ shorter
than $D$, i.e. $M = \cup_{i=0}^{D}E^i$ (note that
$\card(M)=\frac{d^D-1}{d-1}$), and $\varnothing$ is the empty
word. We thus transform sequences as finite sets of
(\emph{context,letter}) couples, where the $context$ is a $D$-long
subsequence of the initial sequence and the $letter$ is the element next to
it. This transformation is justified by the fact
that we consider Markovian models below. An element $X\in\mathcal{X}$
can therefore be written as $X=\{ (x^i,x'^i)\}_{i=1..N_X}$ where
$N_X$ is the cardinal of $X$ (which we will also note $|X|$) and for all $i$, $x^i\in E^D$ and
$x'^i\in E$.
\subsection{Context-tree models}
Context-tree distributions require the definition of a complete
suffix dictionary (c.s.d) $\mathcal{D}$. A c.s.d is a finite set
of words of $M\backslash\{\varnothing\}$ such that any
left-infinite sequence has a suffix in $\mathcal{D}$, but no word
in $\mathcal{D}$ has a suffix in $\mathcal{D}$. We note
$L(\mathcal{D})$ the length of the longest word contained in
$\mathcal{D}$ and $\mathcal{F}_{D}$ the set of c.s.d $\mathcal{D}$
that satisfy $L(\mathcal{D})\leq D$. Once this tree structure is
set, we can define a distribution on $\mathcal{X}$ by attaching
one multinomial distribution $\theta_s \in
\Sigma_d$\footnote{$\Sigma_d$ is the canonical simplex of
dimension $d$, i.e. $\Sigma_d=\{\xi=(\xi_i)_{1\leq i \leq d} :
\xi_i\geq 0, \sum \xi_i=1\}$.} to each word $s$ of a c.s.d
$\mathcal{D}$. Indeed, by denoting $\theta = (\theta_s)_{s \in
\mathcal{D}}$ we define a conditional distribution on
$\mathcal{X}$ by the following equation:
\begin{equation}\label{eqn:ct}
\PP_{\mathcal{D},\theta}(X)=\prod_{i=1}^{N_X}\theta_{\mathcal{D}(x^i)}(x'^i),
\end{equation}
where for any word $m$ in $E^D$, $\mathcal{D}(m)$ is the unique
suffix of $m$ in $\mathcal{D}$.

\begin{figure}
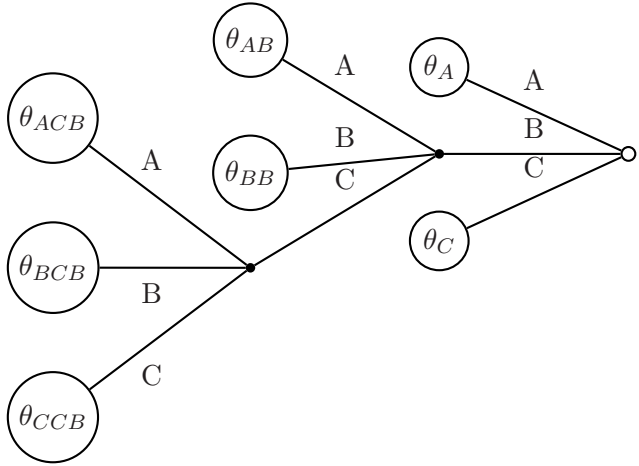

\centering
\pstree[treemode=L,levelsep=*2cm,treenodesize=8pt] {\Tc{3pt}}
    {\pstree{\Tcircle{$\theta_{A}$}^{A}}

     \pstree{\Tdot^{B}}
            {\pstree{\Tcircle{$\theta_{AB}$}^{A}}

            \pstree{\Tcircle{$\theta_{BB}$}^{B}}

             \pstree{\Tdot^{C}}
                {\pstree{\Tcircle{$\theta_{ACB}$}^{A}}{}
                \pstree{\Tcircle{$\theta_{BCB}$}_{B}}{}
                \pstree{\Tcircle{$\theta_{CCB}$}_{C}}{}
                }
            }

     \pstree{\Tcircle{$\theta_C$}^{C}}{}
}
  \caption{Tree representation of a context-tree distribution}\label{fig:ctd}
\end{figure}

We present in Figure \ref{fig:ctd} an example where $E=\{A,B,C\}$,
the maximal depth $D$ is set to $3$ and where
$\mathcal{D}=\{A,AB,BB,ACB,BCB,CCB,C\}$, with corresponding
$\theta_s$ parameters for $s\in\mathcal{D}$, each $\theta_s$ being
a vector of the three-dimensional simplex $\Sigma_3$. We will also
note $\mathcal{P}_D=\{(\mathcal{D},\theta):
\mathcal{D}\in\mathcal{F}_D, \theta\in\Theta_\mathcal{D}\}$ the
set of context-tree distributions of depth $D$.

\subsection{Prior distributions on context-tree models}
\par Having defined a family of distributions $\mathcal{P}_D$ and recalling \eqref{eqn:k2}, we define in this section a prior
probability $\pi(\mathcal{D}, d\theta)$ on $\mathcal{P}_D$. This
probability factorizes as $\pi(\mathcal{D}, d\theta) =
\pi(\mathcal{D})\pi(d\theta | \mathcal{D})$, two terms which are
defined as follows.

\subsubsection{Prior on the tree structure}
\par $\mathcal{F}_{D}$ is the set of complete
trees of depth smaller than $D$. Intuitively it would make sense to put more prior weight on small trees than on large trees. Indeed, the number of different trees with a given number of leaves increases roughly exponentially with the number of leaves. As a result, small trees would have a very low influence compared to big trees if their prior probability was not boosted.
Following \cite{willems95contexttree} we define a simple probability $\pi$ on $\mathcal{F}_{D}$ that has this property by describing a random generation of trees. Starting from the root, the tree generation process follows recursively the following rule: each node has $d$ children with probability $\epsilon$, and $0$ children with probability $1-\epsilon$ (it is then a leaf). In mathematical terms, this defines a branching process. If we denote by $\overset{\circ}{\mathcal{D}}$ the strict suffixes of elements of $\mathcal{D}$, the probability of a tree is given by:
\begin{equation}\label{eq:pi}
\pi(\mathcal{D})=\prod_{s\in\overset{\circ}{\mathcal{D}}}\varepsilon
\prod_{\underset{l(s)<D}{s\in\mathcal{D}}}\left(1-\varepsilon\right)=\varepsilon^{\frac{|\mathcal{D}|-1}{d-1}}\left(1-\varepsilon\right)^{\card\{s\in\mathcal{D}|l(s)<D\}}.
\end{equation}

\subsubsection{Priors on multinomial parameters}
\par For a given tree $\mathcal{D}$ we now define a prior on $\Theta_{\mathcal{D}} = (\Sigma_d)^{\mathcal{D}}$. We assume an independent
prior among multinomials attached to different words with the
following form:
$$
\pi(d\theta|\mathcal{D}) = \prod_{s \in \mathcal{D}}
\omega(d\theta_s) .
$$
Here $\omega$ is a prior distribution on the simplex $\Sigma_d$.
Following \cite{willems95contexttree} a simple choice is to take a
Dirichlet prior of the form:
$$
\omega_{\beta}(d\theta)=\frac{1}{\sqrt{d}}\frac{\Gamma(\sum_{i=1}^d\beta_i)}{\prod_{i=1}^d\Gamma(\beta_i)}\prod_{i=1}^d\theta_i^{\beta_i-1}\lambda(d\theta),
$$
where $\lambda$ is Lebesgue's measure and
$\beta=(\beta_i)_{i=1..d}$ is the parameter of the Dirichlet
distribution. As it has been observed that mixtures of Dirichlet
are a more natural way to model distributions on amino-acids
\cite{brown93using,nips02-LT22} we propose to use such a prior
here. An additive mixture of $n$ Dirichlet distributions is
defined by $n$ Dirichlet parameters $\beta^1,\ldots,\beta^n$ and
by the probabilities $\gamma^1,\ldots,\gamma^n$ of each mixture
(with $\sum_{k=1}^n \gamma^k=1$), and has the following
definition:
\begin{equation}\label{eqn:priormixture}
\omega(d\theta_s) = \sum_{k=1}^n \gamma^k
\omega_{\beta^k}(d\theta_s) .
\end{equation}

\subsection{Triple mixture covariance kernel}
\par Combining the definition of the kernel (\ref{eqn:k2}) with the
definition of the context-tree model distributions (\ref{eqn:ct})
and of the prior on the set of distributions (\ref{eq:pi},
\ref{eqn:priormixture}), we obtain the following expression for
the kernel:
\begin{equation}\label{eq:k}
\KK_\sigma(X,Y) =
\sum_{\mathcal{D}\in\mathcal{F}_D}\pi(\mathcal{D})
\int_{\Theta_{\mathcal{D}}}
\PP_{\mathcal{D},\theta}(X)^{\frac{\sigma}{N_X}}
\PP_{\mathcal{D},\theta}(Y)^{\frac{\sigma}{N_Y}}
\prod_{s\in\mathcal{D}}\left(\sum_{k=1}^n\gamma^k
\omega_{\beta^k}(d\theta_s)\right).
\end{equation}
We observe that~\eqref{eq:k} involves three summations respectively over the trees, the
components of the Dirichlet mixtures, and the multinomial
parameters. This generalizes the double mixture performed in \cite{willems95contexttree} in the context of sequence compression by adding a mixture of Dirichlet, justified by our goal to process protein sequences.

\section{Kernel implementation}\label{sec:implementation}
The definition of the kernel in \eqref{eq:k} does not express a
practical way to compute it. To do so, we propose to adapt the context-tree weighting algorithm, first introduced
in ~\cite{ willems95contexttree}, based on a factorization of
the kernel along the branches of the context-tree. Let us introduce first a few more
notations. We set, given $r \in\mathds{N}$,
$\beta=(\beta_i)_{1\leq i \leq r}\in
\left(\mathds{R}^{+*}\right)^r$ and $\alpha=(\alpha_i)_{1\leq i
\leq r}\in \left(\mathds{R}^{+}\right)^r$:
$$
\mathds{G}_{\beta}(\alpha)\overset{\text{def}}{=} \int_{\Sigma_r}\prod_{i=1}^{r}\theta_i^{\alpha_i}
\omega_\beta(d\theta) = \frac{\Gamma(\beta_\centerdot)}{\prod_{i=1}^{r}\Gamma(\beta_i)}
\frac{\prod_{i=1}^{r}\Gamma(\alpha_i+\beta_i)}{\Gamma(\alpha_\centerdot+\beta_\centerdot)}
,
$$
where $\Gamma$ is the Gamma function \cite{DurEddKroMit98}, $\beta_\centerdot=\sum_{i=1}^r \beta_i$, and
$\alpha_\centerdot=\sum_{i=1}^r \alpha_i$. The quantity
$\mathds{G}_{\beta}(\alpha)$ corresponds to the averaging of
likelihoods $\mathds{Q}_{\theta}(\alpha)$ under a Dirichlet prior
of parameter $\beta$ for $\theta$. We can now divide the algorithm
computation into two phases which can be implemented
alongside.

\subsection{Defining counters} The first step of the
algorithm is to compute, for $e\in E$ and $m\in M$, the following
counters:
$$
\begin{aligned}
\rho_m (X) & = \sum_{i=1}^{N_X}\mathds{1}(x^i=m ),\\
\hat{\theta}_{m,e} (X) & =
\begin{cases}
\frac{\sum_{i=1}^{N_X} \mathds{1}(x^i=m,x'^i=e)}{\rho_m (X)}
 &\text{ if } \rho_m(X)>0\\
\frac{1}{d} &\text{ else}
\end{cases},\\
a_{m,e}(X,Y) & =\frac{\rho_m(X)}{|X|}\hat{\theta}_{m,e}(X) +
\frac{\rho_m(Y)}{|Y|}\hat{\theta}_{m,e} (Y)
\end{aligned}
$$

Counter $\rho_m(X)$ keeps track of the frequency of the motive $m$
in the set $X$ while $\hat{\theta}_{m,e}$ summarizes the empirical
probability of the apparition of letter $e$ after $m$ has been
observed. Finally $a_{m,e}(X,Y)$  takes into account a weighted
average of the transitions encountered both in $X$ and $Y$. The
most efficient way to compute those counters is to start defining
them when $m$ only goes through visited contexts, which are up to
$N_X+N_Y$, and then benefit from the following downward
recursion on the length of the string $m$ when $m$ goes through
all suffixes of visited contexts:
$$
\begin{aligned}
\rho_m(X)&=\sum_{f\in E}\rho_{f.m}(X),\\
\theta_{m,e}(X)&=\frac{\sum_{f\in
E}\rho_{f.m}(X)\theta_{f.m,e}(X)}{\rho_m(X)},\\
a_{m,e}(X,Y)&=\sum_{f\in E}a_{f.m,e}(X,Y).
\end{aligned}
$$
\subsection{Recursive computation of the triple mixture}
We can now attach to each $m$ for which we have calculated the
previous counters the value:
$$
K_m(X,Y) =\sum_{k=1}^{n} \gamma^k
\mathds{G}_{\beta^k}\left(\sigma\left(a_{m,e}(X,Y)\right)_{e\in
E}\right),
$$
which computes two mixtures, the first being continuous on the
possible values of $\theta$ weighted by a Dirichlet prior and the
second being discrete by using the different weighted Dirichlet
distributions given by the mixture $(\gamma^k,\beta^k)$. Here we
assume that the possible values for $\mathds{G}_{\beta^k}$ are
computed beforehand and stored in a table. By defining now the quantity
$\Upsilon_m(X,Y)$, which is also attached to each visited word $m$
and computed recursively:
$$
\Upsilon_m(X,Y)  =
\begin{cases}
K_m(X,Y)& \text{if } l(m)=D \\
(1-\varepsilon)K_m(X,Y)+\varepsilon \prod_{e\in E}
\Upsilon_{e.m}(X,Y) & \text{if } l(m)<D
\end{cases},
$$
we compute the third mixture over the different possible tree
structures of our complete-suffix dictionary by taking into
account the branching probability $\varepsilon$. Indeed, we
finally have, recalling $\varnothing$ is the empty word, that:
\begin{equation}\label{eqn:recurs}
K_\sigma(X,Y) =\Upsilon_\varnothing(X,Y).
\end{equation}
\begin{proof}
In order to prove (\ref{eqn:recurs}), let us first fix a tree $\mathcal{D}$ and observe that, for $X=(x^i,x'^i)_{i=1 \leq N_X}$ and
$Y=(y^i,y'^i)_{1\leq i \leq N_Y}$:
\begin{multline*}
\int_{\Theta_{\mathcal{D}}}
\mathds{Q}_{\mathcal{D},\theta}(X)^{\frac{\sigma}{N_X}}
\mathds{Q}_{\mathcal{D},\theta}(Y)^{\frac{\sigma}{N_Y}}
\prod_{s\in\mathcal{D}}\left(\sum_{k=1}^n \gamma^k \omega_{\beta^k}(d\theta_s)\right)\\
=
\int_{\Theta_{\mathcal{D}}}\prod_{s\in\mathcal{D}}\left(\prod_{e\in
E}\theta_s(e)^{\sigma a_{s,e}(X,Y)}\left(\sum_{k=1}^n\gamma^k  \omega_{\beta^k}(d\theta_s)\right)\right)\\
= \prod_{s\in\mathcal{D}} \sum_{k=1}^n \gamma^k
\int_{\Sigma_{d}}\left(\prod_{e\in
E}\theta_s(e)^{\sigma a_{s,e}(X,Y)}\omega_{\beta^k}(d\theta_s)\right)\\
= \prod_{s\in\mathcal{D}}\sum_{k=1}^n
\gamma^k\mathds{G}_{\beta^{k}}\left(\sigma\left(a_{s,e}\left(X,Y\right)\right)_{e\in
E}\right)= \prod_{s\in\mathcal{D}} K_s(X,Y),
\end{multline*}
 where we have used Fubini's theorem to factorize the integral in
 the second line. Having in mind \eqref{eq:k}, we have thus proved that $\KK_\sigma(X,Y) =
 \sum_{\mathcal{D}\in\mathcal{F}_D}\pi(\mathcal{D})\prod_{s\in\mathcal{D}}K_s(X,Y)$.
 The second part of the proof is identical to the one given in \cite{willems95contexttree}
 ~\cite{stflour} to which we refer to finalize this result.
\end{proof}
The computation of the counters has a linear cost in time and memory with respect to $N_X+N_Y$. As only nodes that correspond to suffixes of $X$ and $Y$ are created, recursive computation of $\Upsilon_m$ is also linear (the values $\Upsilon_m$ on non-existing nodes being equal to $1$). As a result, the computation of the kernel is linear in time and space with respect to $N_X+N_Y$.

\section{Properties}\label{sec:properties}
\par Besides a fast implementation, the kernels (\ref{eq:k}) has several interesting properties that we briefly mention in this section.

\subsection{Sampling, reverse order and mismatches}\label{sec:sampling}
The kernel (\ref{eq:k}) is defined for elements
of $\mathcal{X}$ which are not strings, but sets of pairs $(context,letter)$. We can use this property to allow more flexibility in our kernel by modifying the representation of a sequence as a set of such pairs. We will use throughout this section a toy-example
where we consider $D=2$, $E=\left\{A,B,C\right\}$ and the sequence $ABCA$.
The straightforward representation of this sequence is the set $X=\left\{(AB,C),(BC,A) \right\}$ with $N_X=2$.

In the case where long sequences are compared and one wants to speed up the algorithm, let us first remark that it is possible to represent a sequence by a sub-sampling, typically random, of the original sequence. This would correspond to choosing randomly and uniformly a fixed number of positions in the sequences, and using the pairs $(context, letter)$ at these positions. By the law of large number, the resulting kernel converges to the true value when the number of points sampled tends to infinity, and deviation bounds can be evaluated.

Second, as typical motifs might be found in various directions for
different sequences, one might be tempted to use both the string
and its reverse ordered form, collecting transitions when reading
the protein in both ways. Following our example, this would yield
to $X=\left\{(AB,C),(BC,A),(AC,B),(CB,A) \right\}$.

Third, recent works in protein homology detection have led to
algorithms as well as kernels ~\cite{ leslie03mismatch,smt} that
can handle minor substitutions between $k$-mers\footnote{a $k$-mer
is a substring of length $k$ found in a sequence} to detect
resemblance beyond exact matching of strings. A simple way to
achieve such a tolerance within the framework defined in this
paper is to redefine $\mathcal{X}$ so that transitions stored in a
set $X$ are given a weighting factor. This means that
$\mathcal{X}$ is redefined as $\mathcal{X}=\cup_{n=0}^{\infty}(E^D
\times E \times [0,1])^n$, and similarly an element $X$ of
$\mathcal{X}$ is written as $X=\{ (x^i,x'^i,\mu^i)\}_{i=1..N_X}$
where each $\mu^i$ is in $[0,1]$. Each encountered pair of
context-string can thus be translated into a subset of fixed size
of likely resemblant context-string pairs given a lesser weight in
the overall computation.  Given a $d\times d$ matrix of
letter-to-letter substitution probabilities in amino-sequences,
referring to ~\cite{ blosum} for instance, one can derive a finite
set of weighted neighbors from a simple context string. We note
$p(e|f)$ the probability of $f$ being replaced by $e$ when $e\neq
f$, both $e$ and $f$ in $E$. We also note $p(e|e)=1$. Following
our previous toy example, we can then consider a transition
$(AB,C,1)$ to be split into a set of weighted mismatch transitions
$\left\{(AB,C,1)\right\} \cup \left\{ (ef, C, P(e|A)P(f|B)),
e,f\in E \right\}$ which can be added integrally to $X$ or limited
to the $S$ most likely substitutions. By redefining our notations
to be now
$$
\begin{aligned}
|X|&=\sum_{i=1}^{N_X}\mu^i,\\
\rho_m (X) & = \sum_{i=1}^{N_X} \mu^i \mathds{1}(x^i=m ),\\
\hat{\theta}_{m,e} (X) & =
\begin{cases}
\frac{\sum_{i=1}^{N_X} \mu^i \mathds{1}(x^i=m,x'^i=e)}{\rho_m (X)}
 &\text{ if } \rho_m(X)>0\\
\frac{1}{d} &\text{ else}
\end{cases},
\end{aligned}
$$
the complexity of our algorithm is still linear in $(N_X+N_Y)$ (but with a larger multiplicative constant)
and allows our kernel to take now into account mismatches.

\subsection{Source coding and compression interpretation}
There is a very classical duality between source distributions (a random model to generate infinite sequences) and sequence compression \cite{CovTho91}. Roughly speaking, if a finite sequence $X$ has a probability $P(X)$ of being generated by a source distribution $P$, then one can design a binary code to represent $X$ by $r(X) = -\log_2 P(X)$ bits, up to a 2 bits, using for example arithmetic coding.

When sequences are generated by an unknown source $P$, it is classical to form a coding source distribution by averaging several a priori sources. Under reasonable assumptions, one can design this way universal codes, in the sense that the average length of the codes be almost as short as if $P$ was known and the best source was used. As an example, the context-tree weighting (CTW) algorithm \cite{willems95contexttree} defines a coding probability $P_{\pi}$ for sequences by averaging source distributions defined by context trees as follows:
\begin{equation}\label{eqn:pw}
P_{\pi}(X):=
\sum_{\mathcal{D}\in\mathcal{F}_D}\pi(\mathcal{D})
\int_{\Theta_{\mathcal{D}}}
\PP_{\mathcal{D},\theta}(X)
\prod_{s\in\mathcal{D}}
\omega_{\beta}(d\theta_s),
\end{equation}
where $\omega_{\beta}$ is the Dirichlet prior with parameter $1/2,\ldots,1/2$. Up to the mixture of Dirichlet and the exponents, we therefore see, by comparing (\ref{eqn:pw}) with (\ref{eq:k}), that our covariance kernel between two sequences can roughly be interpreted as the probability under $P_{\pi}$ of the concatenation of the two sequences. In terms of code length, $-\log K(X,Y)$ is roughly to $r_{\pi}(XY)$, the number of bits required by the CTW algorithm to compress $X$ and $Y$ concatenated together.

Suppose now that the kernel is normalized as follows, to ensure $\tilde{K}(X,X)=1$ for any sequence:
$$
\tilde{K}(X,Y)=\frac{\KK(X,Y)}{\sqrt{\KK(X,X)\KK(Y,Y)}}.
$$
We then obtain that $-\log \tilde{K}(X,Y)$ is roughly equal to:
\begin{equation}\label{eqn:red}
r_{\pi}(XY) - \frac{r_{\pi}(XX) + r_{\pi}(YY)}{2}.
\end{equation}
This non-negative number can be interpreted as the difference between the number of bits required to encode $XY$ and the average of the numbers of bits required to encode $XX$ and $YY$.

In spite of its caveats, this derivation gives a useful intuition about the operation performed by the covariance kernel. It also suggests a general approach to derive a kernel for sequences from a compression algorithms, by compressing $XX$, $YY$ and $XY$ successively, and comparing their lengths with (\ref{eqn:red}). Finding conditions on the compression algorithm that ensure that this procedure leads to a valid kernel remain however an open problem that we are currently investigating.

\section{Experiments}\label{sec:experiment}
\par We report preliminary results concerning the performance of the covariance kernel on a benchmark experiment that tests the capacity of SVMs to detect remote homologies between protein domains. This is simulated by recognizing domains that are in the same SCOP\cite{ hubbard97scop} (ver. 1.53) superfamily, but not in the same family, using the procedure described in \cite{ jaakola00discriminative}. We used the files compiled by the authors of ~\cite{ noble02combining}. For each of the 54 families tested, we computed the ROC (Receiving Operator Characteristic) to measure the performance of a SVM based on the covariance kernel (the ROC score is the normalized area under the curve which plots the number
of true positives as a function of false positives). We tested different parameters of our kernel, and compared its performance with the best mismatch kernel presented in \cite{leslie03mismatch}, that performs at a state-of-the-art accuracy level and can be implemented in linear time. The classification was led using the publicly available Gist
2.0.5
implementation of SVM\footnote{\texttt{http://microarray.cpmc.columbia.edu/gist/download.html}}.

Our covariance kernel has several parameters. The depth $D$, the width $\sigma$ and
the branching probability $\varepsilon$ are the most elementary
to play with; the selection of a Dirichlet mixture is a more
difficult choice. Given the large number of parameters and the risk of overfitting the benchmark dataset by carefully optimizing them, we only report preliminary results with two settings. First we used a single Dirichlet distribution with parameters $1/2,\ldots,1/2$, with $D=5$, $\sigma=5$, $\varepsilon=0.5$. Second, we used a basic 3 component Dirichlet mixture that models three classes of amino-acids (hydrophobic/hydrophilic/highly conserved). This mixture, called \texttt{hydro-cons.3comp}, was downloaded from a Dirichlet mixture repository\footnote{\texttt{http://www.cse.ucsc.edu/research/compbio/dirichlets/}}. Other parameters were set to $D=4$, $\sigma=1$ and
$\varepsilon=0.5$.

\begin{figure}[ht]
  \centering
  \includegraphics[angle=270,width=12cm]{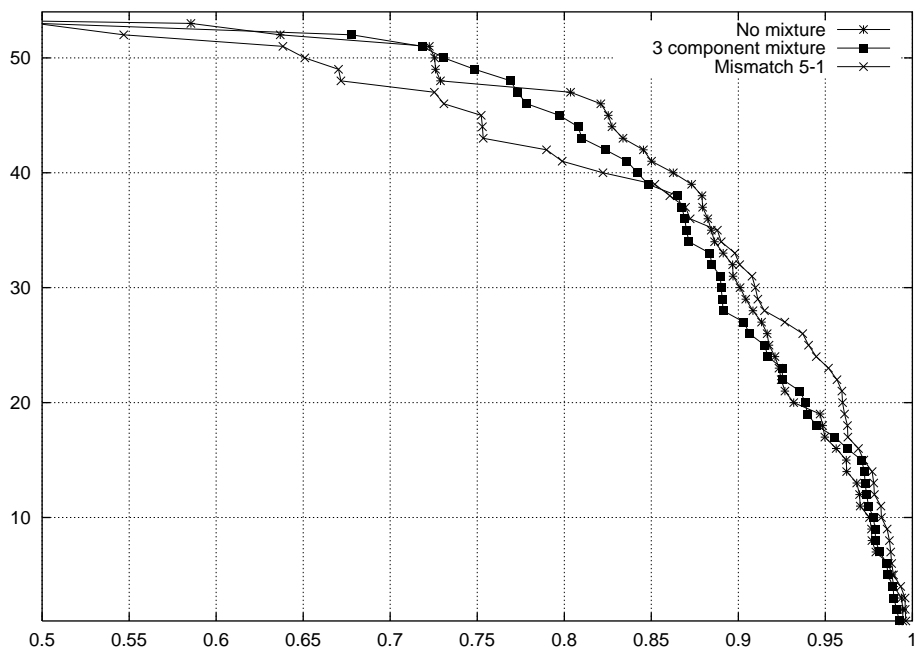}
  \caption{Performance of three kernels on the problem of recognizing domain's superfamily. The curve shows the total number of families for which a given methods exceeds a ROC score threshold.}\label{fig:roc}
\end{figure}

Figure \ref{fig:roc} plots the total number of families for which a given methods exceeds a ROC score threshold. There is no significant difference between the three methods. The mismatch kernel seems to perform better on families with large ROC, while the covariance kernels tend to outperform the mismatch kernel for families with a ROC below $0.85$. This observation is encouraging as it suggests that covariance kernels might be better adapted to difficult problems, corresponding to low sequence similarity, than the mismatch kernel. It is also worth mentioning that we did not test the variant suggested in Section \ref{sec:sampling} to take into account mismatches. The kernel is therefore only based on the same features as the spectrum kernel \cite{leslie02spectrum} which is known to perform worse than the mismatch kernel tested.

\section{Conclusion}
We introduced a novel class of kernels for sequences that are fast to compute and have the flexibility to include prior knowledge through the definition of probabilistic models and prior distribution. The kernel is a covariance kernel based on a family of context-tree models, and makes a link between the string kernels and the theory of universal source coding. On a benchmark experiment of remote homology detection it performs at a state-of-the-art level. Further accuracy improvements are expected from a more careful tuning of the parameters, on the one hand, and from the implementation of sampling strategies to manage mismatches and reverse-ordered similarities, on the other hand.

\section{Acknowledgments}
The authors would like to thank Tatsuya Akutsu, Hiroto Saigo and
Hiroyuki Nakahara for fruitful discussions.
\bibliographystyle{plain}
\small\bibliography{bibliography}
\end{document}